\documentclass[10pt,twocolumn]{article}

\usepackage[T1]{fontenc}
\usepackage[utf8]{inputenc}
\usepackage{newtxtext,newtxmath}
\usepackage[letterpaper,margin=1in,columnsep=0.24in]{geometry}
\usepackage{microtype}
\usepackage{xspace}
\usepackage{amsmath}
\usepackage{booktabs}
\usepackage{multirow}
\usepackage{tabularx}
\usepackage{array}
\usepackage{graphicx}
\usepackage{xcolor}
\usepackage{enumitem}
\usepackage[font=normalsize,labelfont=bf]{caption}
\usepackage{tikz}
\usetikzlibrary{arrows.meta,positioning,fit,backgrounds}
\usepackage[numbers,sort&compress]{natbib}
\usepackage{flushend}
\usepackage[hidelinks]{hyperref}
\usepackage{url}

\definecolor{loreleyblue}{HTML}{2563EB}
\definecolor{loreleygreen}{HTML}{15803D}
\definecolor{loreleyorange}{HTML}{D97706}
\definecolor{ink}{HTML}{0F172A}
\definecolor{muted}{HTML}{475569}
\definecolor{line}{HTML}{CBD5E1}
\hypersetup{
  colorlinks=true,
  linkcolor=loreleyblue,
  citecolor=loreleyblue,
  urlcolor=loreleyblue,
  pdftitle={Loreley: Repository-Scale Program Evolution with Quality-Diversity Search},
  pdfauthor={Mohan Chen}
}
\setlist[itemize]{leftmargin=*,nosep}
\setlist[enumerate]{leftmargin=*,nosep}
\newcolumntype{Y}{>{\raggedright\arraybackslash}X}
\newcolumntype{P}[1]{>{\raggedright\arraybackslash}p{#1}}
\newcommand{\system}{\textsc{Loreley}\xspace}
\newcommand{\code}[1]{\texttt{#1}}

\title{\vspace{-0.6em}\system: Repository-Scale Program Evolution with Quality-Diversity Search}
\author{Mohan Chen\\
\href{mailto:voiletech42@gmail.com}{voiletech42@gmail.com}\\
\href{https://orcid.org/0009-0002-4540-3703}{ORCID: 0009-0002-4540-3703}}
\date{August 2026}

\begin{document}
\maketitle
\vspace{-1.2em}

\begin{abstract}
Sequential agent search accumulates changes from its current champion but
discards alternative branches; independent proposals preserve breadth but
restart from the root. \system instead retains complete repository states in a
Quality-Diversity (QD) archive and samples them as parents or supplies them as
context for later edits. Candidates are Git commits produced in isolated
worktrees and judged by a project-supplied evaluator.

We compare configured Loreley QD, sequential champion editing, and independent
root proposals in a matched Zstandard experiment: seven paired blocks and 48
physical candidate jobs per policy and block (1,008 total), with root-only
initialization and each policy's native concurrency. Validation selected a
winner at each budget checkpoint; an agent-hidden holdout measured the fixed
candidate. At 48 jobs, QD was 0.135\% below Sequential Champion (95\% BCa
interval for the paired effect: $-0.556$\% to $+0.161$\%) and 0.320\% above
Independent Root ($-0.082$\% to $+0.686$\%). Neither contrast established a QD
advantage; Sequential had the highest observed 48-job mean and median.

Archive retention and later sampling did occur. Four of seven final QD winners
had a non-incumbent state in their primary-parent ancestry under a
retrospective one-incumbent rule applied only to the observed QD stream.
Including inspiration edges raised the count to six, without showing that
supplied context caused an edit. Three earlier capability campaigns produced
generation-4, multi-file improvements in two Python libraries and a separate
Zstandard revision. Loreley engaged the intended stepping-stone mechanism, but
the controlled experiment did not show an endpoint benefit at 48 jobs.
\end{abstract}

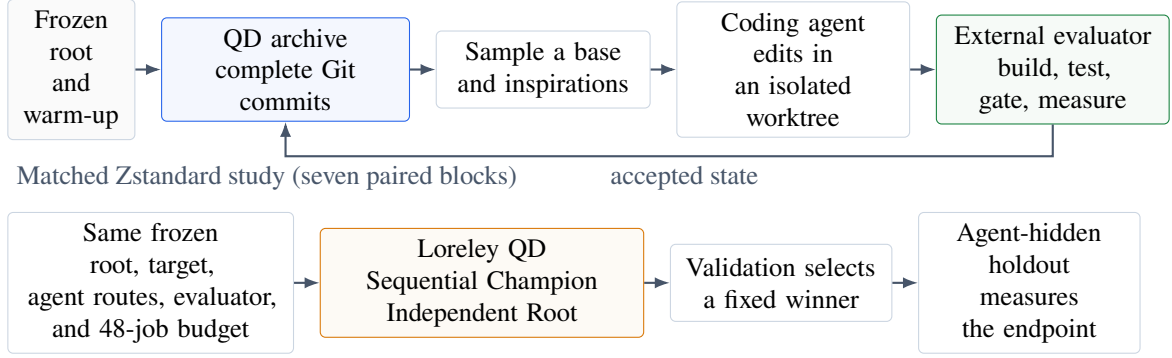
\begin{figure*}[t]
  \centering
  \begin{tikzpicture}[
    node distance=0.72cm and 0.34cm,
    box/.style={draw=line, rounded corners=2pt, align=center,
      minimum height=0.88cm, inner xsep=4pt, inner ysep=4pt, fill=white},
    root/.style={box, text width=1.40cm, fill=black!2},
    archive/.style={box, draw=loreleyblue, fill=loreleyblue!5,
      text width=3.00cm},
    action/.style={box, text width=2.55cm},
    agent/.style={box, text width=2.80cm},
    evidence/.style={box, draw=loreleygreen, fill=loreleygreen!4,
      text width=2.80cm},
    study/.style={box, text width=3.50cm, minimum height=1.02cm},
    policy/.style={box, draw=loreleyorange, fill=loreleyorange!4,
      text width=4.00cm, minimum height=1.02cm},
    outcome/.style={box, text width=2.65cm, minimum height=1.02cm},
    arrow/.style={-{Latex[length=2mm]}, thick, draw=muted},
    note/.style={text=muted, align=left}
  ]
    \node[root] (root) {Frozen root\\and warm-up};
    \node[archive, right=of root] (archive)
      {QD archive\\complete Git commits};
    \node[action, right=of archive] (select)
      {Sample a base\\and inspirations};
    \node[agent, right=of select] (agent)
      {Coding agent edits in\\an isolated worktree};
    \node[evidence, right=of agent] (eval)
      {External evaluator\\build, test, gate, measure};

    \draw[arrow] (root) -- (archive);
    \draw[arrow] (archive) -- (select);
    \draw[arrow] (select) -- (agent);
    \draw[arrow] (agent) -- (eval);
    \draw[arrow] (eval.south) -- ++(0,-0.42cm) -|
      node[pos=0.24,below,note] {accepted state} (archive.south);

    \node[study, anchor=north west] (shared) at
      ([yshift=-0.95cm]root.south west)
      {Same frozen root, target,\\agent routes, evaluator,\\and 48-job budget};
    \node[policy, right=of shared] (policies)
      {Loreley QD\\Sequential Champion\\Independent Root};
    \node[outcome, right=of policies] (validation)
      {Validation selects\\a fixed winner};
    \node[outcome, right=of validation] (holdout)
      {Agent-hidden holdout\\measures the endpoint};

    \draw[arrow] (shared) -- (policies);
    \draw[arrow] (policies) -- (validation);
    \draw[arrow] (validation) -- (holdout);

    \node[anchor=south west, note] at
      ([yshift=0.17cm]root.north west) {Loreley repository-state search};
    \node[anchor=south west, note] at
      ([yshift=0.17cm]shared.north west)
      {Matched Zstandard study (seven paired blocks)};
  \end{tikzpicture}
  \caption{Paper overview. Loreley loops complete Git repository states through
  QD selection, an isolated coding-agent worktree, and external evaluation
  (top). The matched Zstandard study compares three context-selection policies
  under a common 48-job budget; validation fixes each winner before an
  agent-hidden holdout measures it (bottom).}
  \label{fig:overview}
\end{figure*}

\section{Introduction}

Most performance patches in mature software are not standalone programs. They
are repository changes that interact with existing types, tests, build rules,
and public interfaces. A patch can be syntactically valid and still fail to
build, change an output, exceed a resource limit, or disappear under repeated
measurement. Useful repository states are therefore sparse among the states an
editor could produce.

Coding agents can propose coherent changes across files, but the policy that
chooses their next editing context still matters. Sequential Champion search
accumulates edits and concentrates its budget on the best state seen so far.
Independent Root search explores several proposals in parallel but cannot
accumulate improvements across generations. QD offers a third option: retain
several states according to objective trade-offs and repository descriptors,
including states that are not the current champion.

The premise has two testable levels. First, the policy must retain valid
non-incumbents and later sample some as parents or supply them as inspirations.
Second, that continued availability must improve the selected endpoint over
strong simpler policies at a specified budget. The first is an engagement
condition for the intended stepping-stone mechanism, not evidence that the
retained states were useful. A descriptor can separate repositories without
separating useful future edit contexts, and a short search may favor immediate
exploitation by a sequential policy.

\system implements a persistent QD loop over Git repositories. Each candidate
is a commit. A target-specific evaluator is the only authority on correctness
and objective values. Repository embeddings assign valid candidates to a
MAP-Elites grid \citep{mouret2015mapelites}; each cell stores a bounded Pareto
front. Islands may maintain separate archives and supply retained commits as
inspirations. The evaluator can call a local build, a container, a hardware
testbed, or a remote service, so the search core is not tied to Python or to a
particular performance harness. Figure~\ref{fig:overview} summarizes this loop
and the matched evaluation used in the paper.

This paper makes three contributions:

\begin{itemize}
  \item A repository-evolution runtime that combines coding-agent variation,
  isolated Git worktrees, external evaluation, persistent scheduling, and QD
  retention over complete commits.
  \item A matched experiment comparing Loreley QD with Sequential Champion and
  Independent Root under equal attempted candidate jobs. At 48 jobs on
  Zstandard, the experiment does not establish a QD advantage over either
  control.
  \item An empirical separation of policy engagement from endpoint efficacy.
  Lineage and archive records show retention and later sampling of
  non-incumbents, while checkpoint and held-out results show no resulting
  advantage at the tested horizon. Three earlier campaigns provide capability
  cases rather than additional policy comparisons.
\end{itemize}

The matched study compares the configured QD search policy with two simpler
alternatives; it does not isolate the learned descriptor, Pareto retention, or
inspiration sampling individually. We therefore report the negative primary
comparison and the policy-engagement evidence separately.

\section{Repository-Level QD Search}
\label{sec:problem}

We distinguish a \emph{mechanism-engagement condition} from an \emph{efficacy
proposition}. Engagement means that the configured archive keeps several viable
repository states available and that later jobs sample some states that a
one-incumbent rule would not retain when applied to the same observed QD
candidate stream. Efficacy means that this availability improves a held-out
endpoint under a defined budget. Parent and inspiration records test
engagement within the QD stream; only a matched policy comparison tests
comparative benefit. Neither establishes the usefulness of an individual
descriptor cell or retained state.

Let $R_0$ be a fixed root repository and let $c$ denote a Git commit descended
from it. A project evaluator $E$ maps $c$ to either a failure or a successful
record
\begin{equation}
 E(c) = \bigl(\mathbf{f}(c),\, a(c),\, \mathcal{A}(c)\bigr),
\end{equation}
where $\mathbf{f}$ is an ordered objective vector, $a$ is an optional
evaluator-defined candidate identity, and $\mathcal{A}$ contains reports and
artifacts. Correctness, edit-scope, resource, and precision checks are part of
$E$; a fast but invalid commit has no objective vector.

Loreley also computes a behavior descriptor $\mathbf{b}(c)$ from the complete
eligible repository state. After dimensionality reduction, $\mathbf{b}(c)$
selects one cell in a MAP-Elites grid. Within that cell, objective directions
are normalized so that larger is better. Let $\epsilon$ be the configured
objective tolerance. After sorting candidates by commit hash, Loreley skips a
later vector $\mathbf{x}$ when a retained representative $\mathbf{y}$
satisfies $\lvert x_j-y_j\rvert\leq\epsilon$ for every objective $j$.
Among the remaining representatives, $\mathbf{x}$
$\epsilon$-dominates $\mathbf{y}$ when
$x_j\geq y_j-\epsilon$ for every $j$ and
$x_k>y_k+\epsilon$ for at least one $k$. Only non-dominated
representatives remain. When they exceed the configured cell capacity,
crowding distance preserves boundary trade-offs and isolated interior points
\citep{deb2002nsga2}. Cell selection and member selection are separate so that
a cell with a larger Pareto front does not receive more sampling probability by
accident. Specifically, Loreley samples uniformly from occupied cells and then
uniformly from the retained members of the selected cell, after applying batch
exclusions.

The search budget counts physical jobs. A job selects a local base commit and
zero or more inspiration commits, invokes the agents, creates a candidate
commit, and evaluates it. Inspirations are sampled without replacement from
occupied cells within an expanding Chebyshev radius, with a bounded fallback to
other occupied cells. On~a migration step, a non-duplicate elite from another
island replaces the last ordinary inspiration slot. Inspiration supplies
evidence from another lineage; it does not change the Git parent of the new
commit. Parent edges therefore record source ancestry, while inspiration edges
record information supplied to the agent.

Figure~\ref{fig:algorithm} gives the end-to-end policy. Warm-up and ordinary
jobs use the same agent and evaluator; warm-up differs only in that every base
is $R_0$ and no unstable archive coordinate is sampled. A scheduling batch is
drawn from one archive snapshot. Within that batch, base commits and complete
base--inspiration recipes are excluded when alternatives exist. Candidates
that finish during the batch can affect only a later batch.

\begin{figure}[t]
  \begin{enumerate}[leftmargin=1.4em,itemsep=1pt,topsep=2pt]
    \item Evaluate the fixed root and run $w$ root-based warm-up edits. Record
      every outcome; add each valid, artifact-unique warm-up offspring to
      descriptor history.
    \item Fit the learned projection when history reaches $w$. Project the
      eligible warm-up states and admit their objective vectors to per-cell
      Pareto fronts.
    \item Freeze the current archive as the sampling snapshot for the next
      batch. Sample an occupied cell, then a retained member as the base.
    \item Sample distinct nearby retained states as inspiration context,
      expanding the cell radius and then using a bounded global fallback.
    \item Let the planning and coding agents edit the base in an isolated
      worktree. Commit the resulting source and evaluate it externally.
    \item Charge the job budget regardless of outcome. For a valid,
      artifact-unique result, update descriptor history and attempt Pareto
      admission in its cell.
    \item At each refit interval, align the new projection and atomically
      rebuild the archive from the records retained immediately before refit.
      Repeat from step 3 until the physical-job budget is exhausted.
  \end{enumerate}
  \caption{End-to-end Loreley QD policy pseudocode. The formal Zstandard
  treatment instantiates $w=4$, one island, four-job scheduling batches, and
  no migration; Appendix~\ref{app:formal-treatment} gives the remaining
  settings.}
  \label{fig:algorithm}
\end{figure}

\section{System Design}
\label{sec:system}

Figure~\ref{fig:overview} shows the lifecycle implemented by \system. The system
separates target-independent search infrastructure from the target evaluator.
The search infrastructure owns scheduling, agent sessions, worktrees, commits,
embeddings, archive state, lineage, and worker coordination. The evaluator
owns the permitted source scope, build and test commands, workloads, objective
definitions, measurement precision, and acceptance gates.

\subsection{Agent variation over Git commits}

The scheduler samples a base from one island and may attach retained commits as
inspirations. A planning agent receives the frozen campaign goal and
constraints; bounded base-commit history, metrics, evaluation evidence, and key
files; and, for each inspiration, a base-relative trajectory and the same
bounded evidence fields. The inspiration is supplied as context, not checked
out as the worktree. A coding agent then edits the complete base repository in
an isolated Git worktree. Successful work produces exactly one result commit.
The worker verifies the worktree and commit before invoking the evaluator;
agent text is not treated as evidence of correctness or effect.

Every job records the base, inspirations, island, model routes, agent summaries,
usage, timing, and terminal reason. Failed candidates remain in the ledger but
do not enter the ordinary QD archive. Valid candidates that are dominated in
their behavior cell also remain reproducible even though they are not sampled
by default.

\subsection{Repository-state behavior descriptors}

Loreley embeds the eligible files at $R_0$ and incrementally updates the
repository vector for descendant commits. File embeddings are cached by
\emph{(experiment, Git blob SHA)}. A commit vector is the uniform mean over all
eligible file vectors; files excluded by the fixed campaign ignore rules do not
participate. The full vector is stored before projection.

Each island maintains its own PCA history, projection epoch, and grid. During
warm-up, candidates populate the PCA history but are not placed under an
unstable coordinate system. A later PCA refit reprojects retained candidates
and rebuilds the affected archive atomically. This descriptor is a search
coordinate, not a claim that embedding distance measures semantic novelty
perfectly.

\subsection{Pareto cells and islands}

An evaluator may return several objectives with explicit names and directions.
Every objective in the campaign contract must be present and finite before
archive admission. Each cell retains a bounded non-dominated set. Multiple
islands have independent projection and archive state; scheduling proceeds in
round-robin order under one global job cap. At a configurable cadence, a job
may receive a non-duplicate elite from another island as an inspiration. The
base remains local, and the resulting commit remains in the target island.

This design can preserve branches with different descriptors or objective
trade-offs. Section~\ref{sec:results} reports which retained branches were
later sampled and whether the configured policy improved held-out quality.

\subsection{Evaluation and persistent execution}

The evaluator controls source scope, build and test commands, workloads,
objective definitions, precision, and acceptance gates. Loreley stores each
job, commit, metric, embedding, and archive update in PostgreSQL; Redis and
Dramatiq dispatch isolated worker processes. A commit SHA records ancestry, a
tree SHA records exact tracked source, and an optional evaluator identity can
deduplicate measurements of equivalent compiled artifacts. These identities
support correct execution and lineage analysis, but they are not a separate
method claim. Evaluator-equivalent source states share measurements, and only
the first processed representative enters descriptor history and the active
archive; source-distinct duplicates are not retained as independent future
parents.

Search, model, and evaluator concurrency are configured independently. This
allows parallel candidate generation while a benchmark remains serial, or
several calibrated measurement lanes while the policy preserves its own
dependency structure. Failed jobs remain in the run record and consume their
budget slot; only valid candidates can enter the archive.

\section{Experimental Design}
\label{sec:experiments}

The evaluation has two parts. First, a controlled Zstandard study compares
Loreley's QD policy with two simpler policies under matched candidate budgets.
Second, three earlier campaigns show what the configured system produced on two
Python libraries and a separate Zstandard revision. The earlier campaigns are
capability cases, not additional policy replicates.

\subsection{Matched Zstandard policy experiment}
\label{sec:matched-design}

We ran seven paired blocks on one Linux ARM64 host. Each block contained three
configured policies, initialized only from the same frozen root and given 48
attempted physical candidate jobs. The target revision, task instructions,
planning and coding routes, and training evaluator were common. The online
parent and context rules were not: those rules define the policies being
compared. Only the postsearch training rank and validation winner rule were
common. Candidate jobs, rather than successful candidates, dollars, tokens, or
elapsed time, are the primary matching axis.

\begin{table*}[t]
  \centering
  \caption{Online policies in the matched experiment. Warm-up and failed jobs
  count toward the 48-job budget. The comparison retains each policy's native
  adaptivity and concurrency.}
  \label{tab:formal-policies}
  \begin{tabularx}{\textwidth}{@{}P{0.19\textwidth}P{0.25\textwidth}P{0.28\textwidth}Y@{}}
    \toprule
    Policy & Parent & Retained online state & Context and schedule \\
    \midrule
    Loreley QD & Archive member after four root warm-ups & Three-objective fronts in $4^3$ cells & Base plus two inspirations; four-job batches \\
    Sequential Champion & Current training-LCB champion & One incumbent & Incumbent history; serial \\
    Independent Root & Frozen root & None & Root context; four-job batches \\
    \bottomrule
  \end{tabularx}
\end{table*}

The frozen root was \code{c604f825}. Agents could edit Zstandard C sources but
could not access the validation or holdout corpora. Planning used
\code{gpt-5.6-sol}; coding used \code{gpt-5.6-luna}. Loreley QD additionally
used 1,536-dimensional \code{text-embedding-3-small} code embeddings, reduced
to three whitened PCA coordinates. Coordinates were clipped at three standard
deviations, mapped to $[0,1]^3$, and refit after every four artifact-unique
states. Each refit aligned the new projection and rebuilt the archive from the
records retained immediately before refit. The QD arm had one island, no
migration, and frozen depth gates requiring at least 32 post-warm-up jobs, six
archive-parent rounds, generation four, and six distinct archive parents.
Every completed formal QD arm passed these gates; no analyzed block was
excluded or rerun because of a depth-gate failure. Appendix~\ref{app:formal-treatment}
and the machine-readable treatment record give the task contract, warm-up
seeding, sampler radii and fallback, batch snapshot semantics, prompt-context
fields, seeds, and principal generation settings.

The host had 128 logical CPUs and 245 GiB RAM. Training, validation, and
holdout used disjoint generated corpora fixed before formal search. The
evaluator performed a clean release build and upstream checks, then cross-decoded
root and candidate outputs and measured single-thread compression and
decompression at levels 1, 3, and 5 on pinned CPU lanes. Each evaluation began
with 12 alternating root/candidate rounds, one-second warm-ups, and at least
three seconds per level; it extended to 16 rounds only when the predeclared
precision gate failed. The estimator symmetrically trimmed three log effects
per tail. Hard gates covered source scope, tests, release build, round trip,
cross-decode, compressed size, RSS, per-cell throughput, and interval precision.

Within each block we rotated arm start order to reduce calendar-time
confounding. All seven planned blocks enter the analysis, and none was excluded
after score inspection.

We recorded checkpoints at 8, 16, 24, 32, 40, and 48 jobs. Within each
block--policy--checkpoint group, feasible evaluator identities were ranked by
training compression lower confidence bound, and up to ten unique finalists
were measured on validation. The root was always eligible with selector score
1.0. A non-root candidate was eligible only if it passed evaluation, had
compression lower bound strictly above 1.0, decompression geomean at least
0.995, and worst-cell speedup at least 0.98. The highest selector score won;
exact ties preferred the root, then fewer changed lines, then commit and
evaluator identity. The holdout measured that fixed winner and never changed
selection.

The finalist width was fixed after search completion and before validation or
holdout outcomes were observed. A post-hoc replay at the originally planned
width of five changes several selected outputs but not the qualitative primary
conclusion (Appendix~\ref{app:selection-width}).

The primary endpoint is the 48-job holdout compression-throughput ratio. This
axis follows the frozen online training rank and target-specific validation
rule; decompression and worst-cell performance enter as gates rather than the
primary score. Each block gives a paired log difference between QD and one
control. If $d_b$ is the paired log-ratio difference in block $b$, the reported
percent effect is $100\{\exp(7^{-1}\sum_b d_b)-1\}$. We also report a
20,000-resample paired BCa 95\% interval, an exact sign-flip test over all
$2^7=128$ sign assignments,
and Holm correction over the two target-local contrasts. The sign-flip
interpretation assumes symmetry/exchangeability of the paired block
differences; the blocks are separate searches on the same target and host, not
a sample of repositories. The six checkpoint curves are descriptive because
the primary inference was fixed at 48 jobs. At each checkpoint validation
reselects from the candidates available at that time, so holdout winner quality
need not be monotone in job count.

\subsection{Earlier capability campaigns}

We also report three earlier campaigns as capability cases. They used frozen
repository revisions, external evaluators, and human-written seeds that
represented different optimization hypotheses. Seeds were used because the
system did not yet have a reliable way to generate a diverse initial set; they
were evaluated normally and counted toward the job budget. All three reported
candidates are generation-4 descendants rather than selected seeds. These
campaigns do not compare search policies or estimate the effect of seeding.
The abbreviated upstream/root pairs were \code{bff75ed}/\code{97aff4f},
\code{6568072}/\code{4fb992f}, and \code{82d322c}/\code{5b3fe47}, respectively;
the experiment roots add only campaign-control files.

\begin{table*}[t]
  \centering
  \caption{Earlier capability campaigns. Successful outcomes are shown in
  parentheses. ``Post-hoc'' marks the replacement made after the registered
  Pathspec candidate failed its allocation gate.}
  \label{tab:campaigns}
  \begin{tabularx}{\textwidth}{@{}Y r P{0.20\textwidth}r P{0.25\textwidth}@{}}
    \toprule
    Target & Jobs & Candidate & Generation & Selection \\
    \midrule
    \code{markdown-it-py} & 64 (54) & \code{b10adb6} & 4 & Frozen before validation \\
    \code{python-pathspec} & 64 (45) & \code{9d977f0} & 4 & Post-hoc \\
    Zstandard & 220 (211) & \code{fe39bee8} & 4 & Expanded-validation winner \\
    \bottomrule
  \end{tabularx}
\end{table*}

The Python evaluators used deterministic compile, match, or rendering workloads
inside controlled runtimes and enforced upstream tests, output equivalence,
public API, edit-scope, packaging, and allocation gates. The earlier Zstandard
campaign built a release CLI and used paired single-thread compression and
decompression measurements at levels 1, 3, and 5, together with cross-decode,
size, RSS, scope, and precision gates. Agents could not access validation or
holdout inputs. Ratios above one favor the candidate; aggregates are geometric
means unless stated otherwise.

The paper evidence bundle contains the frozen reports, candidate-selection
qualifications, measurement records, and resource accounting for all three
cases. Appendix~\ref{app:capability-records} gives the four-split Zstandard
record because its validation and holdout roles differ; the main results retain
one row per repository.

\section{Results}
\label{sec:results}

\subsection{Matched policy comparison}

The controlled study scheduled 1,008 jobs and produced 948 successful
candidates. Table~\ref{tab:policy-endpoint} shows the final validation-selected
holdout winner in each block. Sequential Champion has the largest endpoint mean
and median; Loreley QD is intermediate; Independent Root is lowest. A
predefined ``useful'' compression gain of at least 0.5\% occurred in seven of
seven Sequential blocks, six of seven QD blocks, and two of seven Independent
blocks.

\begin{table*}[t]
  \centering
  \caption{Matched Zstandard result at 48 candidate jobs per block and arm.
  Values are compression-throughput gains over the root for seven
  validation-selected holdout winners.}
  \label{tab:policy-endpoint}
  \begin{tabularx}{\textwidth}{@{}Y r r P{0.25\textwidth} r@{}}
    \toprule
    Policy & Mean (\%) & Median (\%) & Range (\%) & $\geq 0.5\%$ \\
    \midrule
    Independent Root & $+0.502$ & $+0.412$ & $+0.195$ to $+1.331$ & 2/7 \\
    Loreley QD & $+0.824$ & $+0.739$ & $+0.062$ to $+1.390$ & 6/7 \\
    Sequential Champion & $\mathbf{+0.960}$ & $\mathbf{+0.819}$ & $+0.514$ to $+1.789$ & 7/7 \\
    \bottomrule
  \end{tabularx}
\end{table*}

The paired QD effect was $+0.320$\% relative to Independent Root (95\% BCa
interval $[-0.082,+0.686]$\%; Holm $p=.375$) and $-0.135$\% relative to
Sequential Champion ($[-0.556,+0.161]$\%; Holm $p=.547$). A positive effect
favors QD. Neither paired interval excludes zero. The experiment therefore does not
establish that QD improves final held-out performance over either control. It
also does not establish equivalence: the intervals still allow a modest effect
in either direction. The observed ordering is specific to Zstandard, the two
GPT routes, root-only initialization, and a 48-job horizon.

The small number of pairs is visible in simple sensitivities. Removing one
block at a time moves the QD--Sequential point effect from $-0.226$\% to
$+0.020$\%, while QD--Independent remains positive from $+0.182$\% to
$+0.472$\%. These ranges do not replace the frozen analysis; they show why the
result should not be read as a stable ranking across tasks. An exploratory
Sequential--Independent contrast, outside the frozen primary family, is
$+0.456$\% with exact sign-flip $p=.125$.

Compression was the primary endpoint. A post-hoc check using the same fixed
winners and their combined compression/decompression throughput gives QD
effects of $-0.531$\% relative to Sequential (BCa 95\% interval
$[-2.407,+0.163]$\%) and $+0.367$\% relative to Independent
($[+0.128,+0.552]$\%); unadjusted exact $p$-values are .719 and .047.
Decompression alone gives $-0.926$\% ($[-4.187,+0.324]$\%, $p=.703$) and
$+0.413$\% ($[+0.084,+0.700]$\%, $p=.078$), respectively. The Sequential
intervals are widened by its unusually large decompression gain in Block 6.
These analyses were not part of the primary test family and received no
multiplicity correction; the nominal $p=.047$ is therefore not confirmatory.
Per-block values appear in Appendix~\ref{app:secondary-endpoints}.

Figure~\ref{fig:method-efficacy} shows how the selected winners changed with
budget and exposes the seven endpoint pairs. At 8 and 16 jobs, QD has the
largest median. Sequential overtakes it at 40 and 48 jobs. These are
descriptive checkpoints, not six independent tests, and the curve is not a
  cumulative-best curve: validation reselects among up to ten finalists at each
  checkpoint. The endpoint scatter makes the run-to-run result concrete. QD exceeds
Independent Root in five blocks and Sequential Champion in three.

\begin{figure*}[t]
  \centering
  \includegraphics[width=\textwidth]{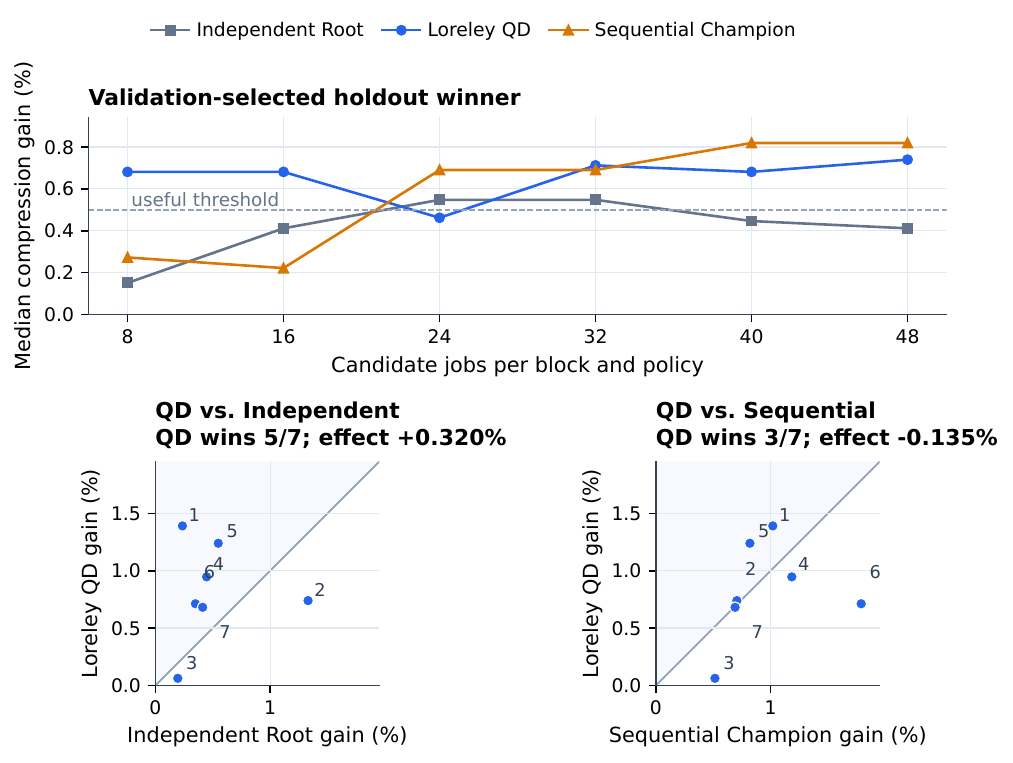}
  \caption{Matched Zstandard policy comparison. Top: median holdout gain of
  the validation-selected winner at each common budget; checkpoint values are
  descriptive. Bottom: the seven paired 48-job endpoints, with
  block numbers and an equal-performance diagonal. Primary inference uses only
  the final paired contrasts in Table~\ref{tab:policy-endpoint}.}
  \label{fig:method-efficacy}
\end{figure*}

\subsection{Archive activity and lineage retention}

The configured QD policy engaged its archive in all seven blocks, and every
final QD winner was a descendant rather than the root. We label a candidate
\emph{non-incumbent at admission} when its training compression lower bound is
strictly below the maximum of the root score 1.0 and all earlier QD candidate
scores; exact ties are not labeled non-incumbent. This is a one-incumbent rule
applied retrospectively to the same observed QD stream. It is not a
counterfactual claim about which candidates the separately run Sequential arm
would have generated.

Four winners had at least one such state in their primary-parent ancestry. Two
more had one only in the larger graph obtained by recursively following parent
and inspiration edges, giving six of seven for that dependency definition.
Across the 67 nodes in those combined graphs, 49 were non-incumbents at
admission and later appeared on an outgoing edge: 13 only as parents, 18 only
as inspirations, and 18 in both roles. An inspiration edge means that the
commit was resupplied as context; it does not show that the agent used its
contents or that it caused the resulting edit.

The four primary-parent lineages contained 15 non-incumbent ancestors. The
selected final descendants occurred 3--42 logical-job ordinals after those
ancestors were admitted. In three of the four lineages, at least one final
descendant exceeded the incumbent score recorded when its ancestor entered the
stream; this held for seven of the 15 ancestors. The remaining cases did not.
This is within-policy evidence of delayed branch survival, not an estimate of
causal benefit or of the best descendant each branch could produce.

Summaries from the seven frozen QD databases give a wider view. A block
used a median of 20 distinct primary parents (range 15--23), with median
entropy-effective support of 17.1 parents. The median parent-revisit lag across
blocks was 5.5 jobs, and the deepest candidate in a block reached generation 5
to 9. Final archives occupied 9 to 12 of 64 cells. These diagnostics show that
QD maintained and revisited several editable contexts; they do not identify
whether the learned cells chose useful contexts, or whether 48 jobs were long
enough for delayed branches to pay off.

Archive retention and later sampling were therefore observed, satisfying the
policy-engagement condition. The efficacy proposition did not receive
corresponding support at 48 jobs. Selected candidate examples are reported in
Appendix~\ref{app:candidate-examples}; they do not replace the paired policy
statistic.

\subsection{Earlier capability campaigns}

Across the three capability cases, 348 terminal jobs yielded 310 successful
outcomes and 38 failures. Table~\ref{tab:results} reports one candidate per
repository. Effects refer to different workloads and are not pooled.

\begin{table*}[t]
  \centering
  \caption{Earlier capability-case results, reported as throughput gain (\%).
  Brackets are 95\% fixed-candidate measurement intervals where available;
  they do not estimate cross-repository generalization. H and F denote the
  original holdout and fresh sealed corpus.}
  \label{tab:results}
  \begin{tabularx}{\textwidth}{@{}P{0.21\textwidth}P{0.15\textwidth}Y Y@{}}
    \toprule
    Target & Candidate & Selection measurement & Additional fixed-candidate measurement \\
    \midrule
    \code{markdown-it-py} & \code{b10adb6} & $+6.750$ [$+6.310,+7.170$] & --- \\
    \code{python-pathspec} & \code{9d977f0} & $+25.140$ & $+24.690$ [$+23.330,+26.060$] \\
    Zstandard & \code{fe39bee8} & $+1.234$ [$+1.156,+1.312$] & H: $+1.173$ [$+1.102,+1.245$]; F: $+0.891$ [$+0.522,+1.261$] \\
    \bottomrule
  \end{tabularx}
\end{table*}

The Markdown winner improved all 28 validation documents, and the Pathspec
winner improved all five reference workloads. All three candidates passed the
target-specific correctness, scope, packaging or build, resource, and
allocation or throughput gates described above. The Zstandard selection-set
interval is not adjusted for selecting among the expanded-validation Top 10;
Appendix~\ref{app:capability-records} distinguishes the split roles.

For \code{markdown-it-py}, the best seed measured +3.23\% on training before
four generations accumulated five-file fast paths in inline HTML matching,
rendering, token attributes, escaping, dispatch, and normalization. The fixed
winner measured +6.99\% on training and +6.75\% on a separate 28-document
validation set. For Pathspec, the reported lineage began with a
$-0.22$\% seed and advanced through four generations to +25.36\% on training.
Its generation-3 parent remained available while 20 other jobs completed before
being sampled again for the final step. The registered training winner failed a
fixed absolute allocation gate at the reference shape; the reported candidate
was selected afterward, so its later replication does not remove the post-hoc
selection status.

\subsection{Earlier Zstandard split results}
\label{sec:zstd-results}

Expanded validation selected training-rank-10 candidate \code{fe39bee8}, an
evolved generation-4 commit rather than a manual seed. Its compression
throughput gain was $+1.234$\% on that selection set, $+1.173$\% on the
original holdout, and $+0.891$\% on a newly generated sealed corpus. The
patch combines a zero-literal fast path, a compression hot-path change, and an
eight-byte histogram update unroll across three files. Compressed size was
unchanged on every measured split and peak RSS rose by 0.031 MiB.

Appendix~\ref{app:capability-records} reports all four split measurements and
their protocol status. In brief, the candidate's original-holdout score was
unknown when expanded validation selected it, although that corpus had already
been opened for the initial Top-3 winner. The fresh corpus was sealed before
measurement, but its deterministic recipe was chosen after candidate fixation.
These are complementary fixed-candidate checks, not an untouched
study-level confirmation or a selection-adjusted estimate.

\section{Discussion}

The matched study separates two questions that are easy to conflate. Loreley
did retain and later sample alternative repository states. That engagement did not
produce a better held-out endpoint than Sequential Champion at 48 jobs. The
following observations constrain, but do not settle, why.

\subsection{Sequential search at short budgets}

The three capability-case winners were all fourth-generation commits, but
multi-generation editing is not unique to QD. Sequential Champion produced a
generation-5 best example in the matched study, while final QD winner depth had
median three and range one to seven. Sequential search can therefore accumulate
substantial changes without maintaining several live branches. Loreley's extra
mechanism is the continued availability of several trajectories; that creates
an opportunity cost unless a retained alternative later yields a better edit.

\subsection{The 48-job horizon}

QD can, in principle, retain stepping stones whose immediate objective value
would not keep them in a single-incumbent search. A runtime analysis proves
such an advantage for MAP-Elites on two particular combinatorial problem
classes \citep{qian2024qdhelpful}; it does not imply the same advantage for
repository optimization. Search horizon is nevertheless a material open
variable here. SATLUTION reports 70 solver revisions, each evaluated on 400 SAT
instances, or about 28,000 solver-instance executions
\citep{yu2025satlution}. This provides scale context, but the units are not
normalized: a Loreley job proposes one repository candidate and performs
repeated multi-cell measurements, while a SATLUTION execution runs one solver
on one instance. We therefore do not treat the counts as a compute, cost, or
sample-efficiency ratio.

The matched experiment supplies seven separate search replicates but only 48 candidate jobs
per policy and block. Its checkpoint curve does not show a late QD advantage:
Sequential Champion catches QD by the final two checkpoints. Still, the QD
mechanism is designed precisely for branches whose payoff arrives after more
than one edit, and the lineages show that such branches were retained and
sampled again. The
present data cannot distinguish an insufficient horizon from an ineffective
descriptor, archive, or sampling rule. More budget may expose a later
advantage, add no benefit, or amplify the same representation limits. Within
the measured horizon, Sequential has the highest observed 48-job mean and
median.

\subsection{Archive and descriptor behavior}

Table~\ref{tab:archives} reports the final archive snapshots. An occupied unit
is an island-cell pair, because the Python islands had independent PCA models
and archives. Final coverage was between 14.1\% and 17.2\%, but the denominators
and projections differ, so these percentages are not a cross-task performance
ranking.

\begin{table}[t]
  \centering
  \caption{Final QD archives. The matched-study row is the median of seven
  separate QD blocks; its range is given in the text.}
  \label{tab:archives}
  \begin{tabularx}{\columnwidth}{@{}Y r r r r@{}}
    \toprule
    Target & Cells & Entries & Occupied & Coverage \\
    \midrule
    markdown-it & 128 & 36 & 18 & 14.1\% \\
    pathspec & 128 & 28 & 19 & 14.8\% \\
    Earlier Zstandard & 64 & 13 & 11 & 17.2\% \\
    Matched Zstandard QD & 64 & 14 & 10 & 15.6\% \\
    \bottomrule
  \end{tabularx}
\end{table}

The earlier Zstandard campaign used seven warm-up states, refit PCA after every
eight new states, clipped whitened coordinates at three standard deviations,
allowed eight Pareto members per cell, and used objective epsilon 0.003. Its final 13 entries
occupied nine singleton cells and two cells with two members; the capacity and
crowding rule were not stressed in the final snapshot. This sparse front can
also reflect the objective epsilon, correlations among the three objectives,
and the sampled candidates; the campaign did not isolate those factors from
the descriptor. The PCA history contained 167 artifact-unique repository-state
representatives; evaluator-equivalent proposals were filtered before history
insertion. Under the final projection, all 167 representatives occupied 13
cells, while the retained archive occupied 11. The first three whitened
components explained 71.0\% of centered variance.

The unprojected repository vectors were close: median pairwise cosine distance
was $3.22\times10^{-6}$ and the maximum was $8.32\times10^{-6}$. This follows
from uniformly averaging file embeddings when a patch changes only a small
part of the repository. The small raw scale does not by itself make the
centered differences noise or show that the candidate distribution is globally
low-dimensional. In this campaign the centered spectrum had participation
ratio 5.31 and entropy effective rank 7.80. The observed coordinates are best
read as geometry of source proposals shaped by changed files and lineages; the
experiment did not test whether they correspond to distinct mechanisms or
runtime behavior. Of 51 candidates admitted
during the run, 22 (43.1\%) would occupy a different cell under the final PCA
than under their admission projection. This is cumulative admission-to-final
churn, not churn at every refit. Change-weighted or diff-restricted descriptors
are direct alternatives for a controlled comparison. We do not report a
post-hoc scalar QD score because the three objectives had no preregistered
common normalization. In the matched study, final QD archives contained 11--17
entries and occupied 9--12 cells. That was enough to support many parents, but
not enough to beat Sequential at the primary endpoint. Coverage alone is
therefore not evidence that the descriptor separates candidates by future
mutation value.

\subsection{Short measurements screened a narrow frontier}

Zstandard used four- and eight-round training measurements to screen a wide
frontier. Among the training Top 10, compression lower bounds spanned only
0.276 percentage points, and the median distance from a point estimate to its
lower bound was 0.541 percentage points. Fixed eight-round validation reduced
that median distance to 0.129 percentage points, yet the validation winner had
ranked tenth on training.

A calibrated four-lane root/root experiment measured maximum aggregate log
bias of 0.000314, or about 0.031\%. Paired execution controlled common host
effects, but the finalist intervals still overlapped. The training measurements
were useful for rejecting large regressions and finding a positive band; they
did not provide a stable fine ordering inside the selected Top 10.

\subsection{Time, model usage, and failures}

Table~\ref{tab:resources} keeps jobs, successful outcomes, time records, and
dollar semantics separate. The Python amounts are request-level estimates from
the campaign proxy; the earlier Zstandard amount is a model-catalog estimate.
The formal study's ledger attributes generation and embedding costs to each
arm. None of these values is a provider invoice, and local compute is unpriced.

\begin{table*}[t]
  \centering
  \caption{Recorded resources. Time semantics differ: capability cases record
  elapsed or active runner time; matched-study rows record cumulative worker-wave
  time and overlap in the experiment calendar. Dollar values are estimates or
  ledger attributions, not provider invoices.}
  \label{tab:resources}
  \begin{tabularx}{\textwidth}{@{}Y r r r r l@{}}
    \toprule
    Target or arm & Jobs & Success & Time (h) & USD & Basis \\
    \midrule
    markdown-it capability & 64 & 54 & 4.35 & \$2.08 & Proxy \\
    Pathspec capability & 64 & 45 & 3.91 & \$2.49 & Proxy \\
    Earlier Zstandard capability & 220 & 211 & 5.31 & \$60.25 & Catalog \\
    Matched Independent Root & 336 & 316 & 92.25 & \$185.30 & Attributed \\
    Matched Loreley QD & 336 & 313 & 32.00 & \$140.05 & Attributed \\
    Matched Sequential Champion & 336 & 319 & 85.68 & \$192.54 & Attributed \\
    \bottomrule
  \end{tabularx}
\end{table*}

The matched study cost \$517.89 in attributable generation and embedding
records. Equal jobs did not mean equal spend: QD used more usage events but a
lower recorded generation cost, while policy dependency and batching produced
very different worker-wave times. These are secondary resource outcomes, not
alternative denominators for the frozen primary comparison. Across the matched
study 60 jobs failed; across the capability cases 38 failed. A failed or no-op
job still consumed its physical job slot, preserving the comparison between
policies as actually run.

\section{Related Work}

Genetic improvement has long used search to modify existing software for
runtime, memory, energy, repair, and functionality objectives
\citep{petke2018gi}. Within QD, MAP-Elites discretizes a behavior space
\citep{mouret2015mapelites}; MOME places a Pareto front in each cell
\citep{pierrot2022mome}; and AURORA periodically relearns an unsupervised
descriptor \citep{cully2019aurora}. Loreley's per-cell fronts and learned
coordinates build on these components. Its method contribution is their use in
a persistent coding-agent runtime whose search states are buildable Git
repositories, not any one archive component in isolation.

ELM combined an LLM program mutator with MAP-Elites
\citep{lehman2022elm}. FunSearch joined language-model sampling, executable
evaluation, and a program database \citep{romeraparedes2024funsearch};
AlphaEvolve broadened the editable program and evaluator set
\citep{novikov2025alphaevolve}. Open-source CodeEvolve adds island-based
CVT-MAP-Elites, inspirations, and refinement on algorithmic tasks
\citep{assumpcao2025codeevolve}. QDEvo likewise combines LLM variation,
multiobjective QD, and pretrained code embeddings, but searches algorithmic
heuristics rather than executable repositories \citep{khanh2026qdevo}.

RHO's HELIX optimizer is a direct repository-level antecedent
\citep{elmaaroufi2026rho}. It treats a multi-file policy repository as the
evolutionary artifact, uses a tool-enabled coding agent in isolated worktrees,
scores with a user-supplied environment, and retains several repositories on a
per-instance coverage frontier. Loreley instead discretizes learned
whole-repository descriptors and keeps a bounded multiobjective Pareto front in
each cell. The present paper also compares its configured policy with
sequential and independent repository search, showing that observed branch
retention did not imply a 48-job endpoint advantage.

CktEvo is the closest repository-level MAP-Elites comparison
\citep{shi2026cktevo}. It evolves multi-file Verilog repositories with
hand-designed RTL descriptors, island migration, a synthesis/formal evaluator,
and one scalar area-delay elite per bin. Loreley's evaluator is target-supplied
and its cells retain multiobjective fronts. Our claim is therefore not priority
for repository-level evolution, agent mutation, learned QD descriptors, or
per-cell Pareto retention; it is the configured repository-search method and
its matched mechanism-versus-endpoint study.

Other systems evolve or coordinate repository states. EvoGit records agent
collaboration as a Git phylogeny \citep{huang2025evogit}; SATLUTION evolves
C/C++ SAT solvers \citep{yu2025satlution}; ABCEvo and HORIZON target hardware
tool and project repositories \citep{yu2026abcevo,yu2026horizon}; and GEAR
maintains complete machine-learning research states \citep{jeddi2026gear}.
Vesper studies harness isolation and evaluator exploits
\citep{ishibashi2026vesper}.

Controlled studies of AI-research search policies reach a related empirical
question. Heuresis holds a research framework fixed while comparing greedy,
MAP-Elites, Islands, Go-Explore, and divergent strategies across quality,
diversity, and novelty \citep{antoniades2026heuresis}. FML-bench separates
search strategy from execution infrastructure across 18 machine-learning
tasks; its simple greedy hill climber nearly matches the best tree-search
agent, and diversity alone is not associated with final performance
\citep{zou2026fmlbench}. Loreley studies the same tension between concentrated
and broader search in buildable multi-file software repositories, with Git
commits as persistent states and held-out paired-block endpoints. Simple
independent or sequential baselines can also match more elaborate code-evolution
methods on some tasks \citep{gideoni2026baselines}, and EvoTrace shows that
several edit mechanisms can produce score gains \citep{pelleriti2026evotrace}.
These findings motivate the strong policy controls and restrained attribution
used here.

\section[Limitations and Next Experiments]
        {Limitations and\\Next Experiments}
\label{sec:limitations}

The matched comparison covers one repository, one pair of GPT routes, one
Linux ARM64 host, and seven blocks. Its confidence intervals are substantially
more informative than a single run but remain compatible with modest effects
in either direction. The 48-job horizon is especially important: QD retained
delayed branches, yet the experiment may have ended before such branches could
compound. That explanation is plausible, not established. The same data are
also compatible with a descriptor or archive that preserves alternatives
without improving their expected payoff.

The paired analysis estimates variation across repeated searches on one fixed
task. It does not estimate cross-repository variation. Candidate timing
intervals, validation selection uncertainty, repeated-search variation, and
cross-task generalization are different layers; the block-level BCa interval
addresses only the third.

The experiment compares one configured Loreley QD policy with two configured
controls. It does not identify the separate effects of the learned descriptor,
Pareto retention, inspiration sampling, or concurrency. Archive coverage and
parent entropy show that the archive was sampled; they do not show that cell
distance predicts mutation utility. A component study should compare the
current repository-mean embedding with diff-restricted, structural, and random
coordinates under the same longer-horizon policy protocol.

The descriptor is non-stationary. A PCA refit rebuilds the archive from
currently retained records; candidates discarded under an earlier projection
are not reconsidered when coordinates change. Archive membership is therefore
path-dependent on the projection schedule. We did not perform a full-history
or fixed-basis replay for the formal blocks, so the parent and dependency
diagnostics are conditional on the implemented retained-only rebuild rule.

Root-only initialization removes manual seed quality from the matched study.
The three capability cases did use human-written seeds because no reliable
automatic procedure was available for generating diverse starting hypotheses.
Their reported candidates were evolved descendants rather than direct seed
selections, but those cases do not estimate the effect of initialization.

The capability cases retain their original selection qualifications.
\code{markdown-it-py} was frozen before separate validation. Pathspec is a
post-hoc replacement followed by fixed-candidate replication. The earlier
Zstandard candidate was selected on expanded validation; its own original
holdout score was unseen at selection although the corpus had been opened for
another candidate, and the fresh-corpus recipe was chosen after candidate
fixation. These cases establish capability, not a shared inferential claim.

The workloads measure throughput on two Python libraries and one C compressor;
they do not measure maintainability, upstream acceptance, production traffic,
or broader software-engineering objectives. The next decisive experiment is a
longer matched curve on Zstandard together with a second repository using the
same three policies. Such a study would test whether QD's retained branches
eventually repay their early opportunity cost or whether Sequential remains
the better policy in this regime.

\section{Artifacts and Reproducibility}

Loreley is released under Apache-2.0 at
\url{https://github.com/NeapolitanIcecream/loreley}. The repository contains
the system source, case-study reports, published candidate patches, and
machine-readable evidence. For the matched study,
\code{zstd\_formal\_records.json} contains the finalist groups, validation
measurements, fixed-winner holdout measurements, and parent/inspiration graphs
needed for the paper's results. \code{zstd\_formal\_treatment.json} records the
task, agent, warm-up, descriptor, archive, sampler, and batch settings described
in Appendix~\ref{app:formal-treatment}. The validator replays the winner rule and
the validation-to-holdout mapping, then recomputes the primary and secondary
contrasts, sensitivity analyses, and public lineage counts from full-precision
records. A separate package gate verifies an explicit set of paper-critical
files and derives raw-record requirements from the included evidence JSON, so
the distributed review package must contain every cited capability result and
the earlier Zstandard preregistration at the recorded hash.

The public record omits prompts, private paths, hidden corpus contents, and
candidate source from the formal study. Database-only archive diagnostics are
reported as descriptive aggregates and are not claimed to be externally
reconstructable. The \path{paper/evidence/} README states these boundaries and
gives the commands used to regenerate figures and validate the released
records.

\section{Conclusion}

Loreley applies QD search to complete repository states. The archive kept
non-incumbent commits available; later jobs sampled some as parents or
resupplied them as inspiration context. Three capability campaigns also show
that the system can produce cumulative, multi-file improvements.

The controlled result is narrower. On one Zstandard revision and a 48-job
horizon, Loreley QD did not beat Sequential Champion, and its positive point
estimate over Independent Root remained uncertain. Archive retention and later
sampling were observed, but the experiment did not establish an endpoint
benefit. Longer
horizons, a second repository, and component controls are needed to determine
whether different descriptors or retention policies can turn that activity
into an advantage over a strong sequential baseline.

\section*{Declarations}

\paragraph{Funding.}
This work received no external funding.

\paragraph{Generative AI and tool use.}
Generative AI systems had two roles in this work. The model routes used as
experimental components are reported in the methods and released artifacts.
Separately, OpenAI Codex assisted with software development, experiment
planning and orchestration, literature research, evidence analysis, figure
preparation, and drafting and revising the manuscript. ChatGPT and Claude Opus
5 were used for manuscript review. The author made the final decisions on the
study design, analysis, interpretation, claims, and wording; reviewed the
resulting artifacts and manuscript; and accepts full responsibility for the
content. All figures were generated programmatically from released records; no
generative-image model was used.

\paragraph{License.}
\textcopyright\ 2026 Mohan Chen. This work is licensed under the
\href{https://creativecommons.org/licenses/by/4.0/}{Creative Commons
Attribution 4.0 International License}.

\appendix

\section{Formal Zstandard treatment}
\label{app:formal-treatment}

The settings below instantiate Figure~\ref{fig:algorithm} in the matched
experiment. They are treatment definitions, not settings chosen after observing
the endpoints. The two controls share the target, agents, evaluator, and
postsearch winner rule but use the online policies in
Table~\ref{tab:formal-policies}. The task and ignore texts are hash-bound in
\code{zstd\_formal\_treatment.json}; their literal contents are released in
\path{tools/method_efficacy_experiment/zstd_target.py}.

\begin{description}[style=nextline,leftmargin=0pt,labelindent=0pt,itemsep=0.5em]
  \item[Target contract.] Root \code{c604f825}; edit Zstandard \code{lib} C/H
    sources; preserve upstream tests, round trip and cross-decode, per-cell size
    within 0.1\%, peak RSS within 16 MiB, portability, and the no-special-casing
    rule. Training uses levels 1, 3, and 5.

  \item[Agent runtime.] Kilo 7.4.16 over OpenAI Responses, \code{max} variant;
    planning \code{gpt-5.6-sol} (3,600 s), coding \code{gpt-5.6-luna} (5,400 s),
    whole job 18,000 s, and response-chunk inactivity 900 s. No direct
    temperature or top-$p$ override was recorded.

  \item[Agent context.] Frozen goal, constraints, acceptance criteria, worker
    contract, and sampler facts; bounded base history, change and evaluation
    summaries, up to four metrics, evaluator evidence, and eight key files.
    Each inspiration adds the same bounded evidence and a base-relative
    trajectory. Inspirations are text context; only the base is checked out.

  \item[Warm-up and batching.] Four budgeted root-only warm-up jobs. Valid
    artifact-unique offspring enter descriptor history; the first fit projects
    and considers all eligible warm-up offspring for admission. Ordinary jobs
    are scheduled four at a time from one frozen archive snapshot. Same-batch
    completions become visible only to later batches.

  \item[Descriptor.] Uniform mean of eligible-file
    \code{text-embedding-3-}\allowbreak\code{small} vectors (1,536 dimensions);
    three-dimensional
    whitened PCA with random state 0; coordinates clipped at $\pm3$ standard
    deviations and mapped to $[0,1]$; minimum fit and warm-up of four, history
    capacity 4,096, and refit every four artifact-unique states. Refit aligns
    the projection and rebuilds from records retained immediately before refit.

  \item[Archive.] One island, no migration; $4^3=64$ cells; objectives are
    compression LCB, decompression LCB, and worst-cell speedup; Pareto capacity
    8 and epsilon 0.003. Commit-hash order selects representatives of
    epsilon-equivalent vectors; epsilon-dominance removes dominated
    representatives before crowding-distance overflow. Sampling is uniform
    first over occupied cells and then over retained members.

  \item[Inspirations.] Two distinct retained states per ordinary job;
    expanding Chebyshev radius 1--3, then a global fallback sample of at most
    eight; at most 32 resampling attempts; complete base--inspiration recipes
    have a 64-job cooldown.

  \item[Repetition.] 48 physical jobs per block and at most four unfinished QD
    jobs. The QD base-sampler seed is 20260821. Block $b$ uses
    \code{arm\_seed + 100000(b-1)}; each one-job control campaign additionally
    adds \code{job\_ordinal-1}. All other configured treatment settings are
    fixed across blocks.
\end{description}

\section{Capability-case split records}
\label{app:capability-records}

Table~\ref{tab:zstd-splits} reports the earlier Zstandard candidate on every
measured split. The expanded-validation interval is not adjusted for selecting
the candidate among the training Top 10 on that split. The original-holdout
score was unknown at selection, although its corpus had been opened earlier
for the initial Top-3 winner. The fresh recipe was chosen after candidate
fixation and sealed before measurement. These distinctions affect the
inferential role of each row, not the fixed-candidate measurements themselves.

\begin{table*}[t]
  \centering
  \caption{Candidate \code{fe39bee8} on every measured Zstandard split. Values
  are throughput gains over the root (\%) with 95\% fixed-candidate measurement
  intervals. Worst cell is the minimum gain over levels 1, 3, and 5 in both
  directions.}
  \label{tab:zstd-splits}
  {\setlength{\tabcolsep}{3pt}
  \begin{tabularx}{\textwidth}{@{}P{0.16\textwidth}P{0.19\textwidth}P{0.19\textwidth}P{0.19\textwidth}r r@{}}
    \toprule
    Split & Compression & Decompression & Combined & Worst & Rounds \\
    \midrule
    Training (rank 10) & $+1.145$ [$+0.708,+1.584$] & $+0.002$ [$-0.301,+0.306$] & $+0.572$ [$+0.231,+0.914$] & $-0.357$ & 4 \\
    Expanded validation & $+1.234$ [$+1.156,+1.312$] & $+0.151$ [$-0.008,+0.311$] & $+0.691$ [$+0.592,+0.790$] & $+0.131$ & 8 \\
    Fresh corpus & $+0.891$ [$+0.522,+1.261$] & $-0.170$ [$-0.529,+0.191$] & $+0.359$ [$+0.004,+0.716$] & $-0.182$ & 12 \\
    Original holdout & $+1.173$ [$+1.102,+1.245$] & $+0.017$ [$-0.150,+0.185$] & $+0.594$ [$+0.486,+0.701$] & $-0.146$ & 12 \\
    \bottomrule
  \end{tabularx}}
\end{table*}

The initial Top-3 protocol selected manual seed \code{7b9aef38}, whose sealed
holdout compression effect was +1.019\% (95\% interval +0.962 to +1.076\%).
That registered result remains part of the protocol history. The later
Top-10 validation selected \code{fe39bee8}; reporting this fixed candidate on
each split makes the candidate-level record consistent across the paper while
preserving the original Top-3 conclusion.

\section{Finalist-width sensitivity}
\label{app:selection-width}

The primary analysis selected from up to ten evaluator-unique finalists per
checkpoint. Replaying selection at the originally planned width of five
changed eight of 21 endpoint outputs: six groups selected a different candidate
and two used the root fallback because no Top-5 finalist passed validation.
Five newly selected candidates required fixed-candidate holdout measurement
after the formal holdout had been opened, so this is a post-hoc sensitivity,
not a replacement analysis. The replay gives QD--Sequential $-0.135$\% (BCa
95\% interval $[-0.553,+0.160]$\%, exact $p=.563$) and QD--Independent
$+0.223$\% ($[-0.504,+0.657]$\%, $p=.516$). Both Holm-adjusted $p$-values are
1.0; the qualitative primary conclusion is unchanged.

\section{Selected formal-study candidates}
\label{app:candidate-examples}

The best validation-selected endpoint from each policy illustrates the kinds
of edits produced, but these examples are not a replacement for the paired
statistic. Independent Root found a generation-1 candidate with a $+1.331$\%
compression gain by fusing sequence-code construction with histogram
accumulation; decompression changed by $-0.035$\%. The best QD example reached
$+1.390$\% at generation 4 along a primary-parent lineage
containing retained non-champions, combining a two-word match counter with
zero-literal and sequence-recording fast paths; decompression changed by
$-0.171$\%. Sequential Champion reached $+1.789$\% at generation 5
by accumulating compile-time specialization and a peeled decompression
sequence loop. Its decompression and combined-throughput gains were
$+7.902$\% and $+4.801$\%.

\section{Secondary endpoint records}
\label{app:secondary-endpoints}

\begin{center}
  \captionof{table}{Secondary holdout gains (\%) for the same
  validation-selected 48-job winners as Table~\ref{tab:policy-endpoint}.
  These endpoints were not used to select candidates or expand the primary
  test family.}
  \label{tab:secondary-endpoints}
  {\setlength{\tabcolsep}{3pt}
  \begin{tabular}{@{}r rrr@{}}
    \toprule
    \multicolumn{4}{c}{Decompression} \\
    \cmidrule(lr){1-4}
    Block & Independent & QD & Sequential \\
    \midrule
    1 & $+0.005$ & $-0.171$ & $+0.582$ \\
    2 & $-0.035$ & $+0.737$ & $-0.293$ \\
    3 & $-0.083$ & $-0.226$ & $-0.544$ \\
    4 & $-0.036$ & $+0.253$ & $-0.466$ \\
    5 & $-0.784$ & $-0.017$ & $-0.055$ \\
    6 & $-0.678$ & $+0.259$ & $+7.902$ \\
    7 & $-0.216$ & $+0.219$ & $+0.741$ \\
    \bottomrule
  \end{tabular}}

  \vspace{0.6em}

  {\setlength{\tabcolsep}{3pt}
  \begin{tabular}{@{}r rrr@{}}
    \toprule
    \multicolumn{4}{c}{Combined throughput} \\
    \cmidrule(lr){1-4}
    Block & Independent & QD & Sequential \\
    \midrule
    1 & $+0.120$ & $+0.607$ & $+0.800$ \\
    2 & $+0.645$ & $+0.738$ & $+0.205$ \\
    3 & $+0.056$ & $-0.082$ & $-0.016$ \\
    4 & $+0.205$ & $+0.599$ & $+0.356$ \\
    5 & $-0.120$ & $+0.609$ & $+0.381$ \\
    6 & $-0.166$ & $+0.485$ & $+4.801$ \\
    7 & $+0.097$ & $+0.450$ & $+0.715$ \\
    \bottomrule
  \end{tabular}}
\end{center}

\clearpage
\setlength{\bibsep}{0pt}
\bibliographystyle{plainnat}
\bibliography{references}

\end{document}